\documentclass[a4paper,12pt]{article}
\usepackage{amsfonts}
\usepackage{latexsym}
\usepackage{amsmath}
\usepackage{amssymb}
\usepackage{amssymb}
\usepackage{slashed}
\usepackage{upgreek}
\usepackage{mathrsfs}
\usepackage{bbm}

\usepackage[toc,page]{appendix}

\usepackage{amsthm}

\theoremstyle{remark}

\usepackage[utf8]{inputenc}
\usepackage[english]{babel}

\theoremstyle{definition}

\allowdisplaybreaks

\usepackage{relsize}
\usepackage{graphicx}
\usepackage{comment}

\renewcommand{\thefootnote}{\fnsymbol{footnote}}

\def\appendix#1{\addtocounter{section}{1}\setcounter{equation}{0}
	\renewcommand{\thesection}{\Alph{section}}
	\section*{Appendix \thesection\protect\indent \parbox[t]{11.15cm}{#1}}
	\addcontentsline{toc}{section}{Appendix \thesection\ \ \ #1}}

\def\bbe{{\bf{e}}}

\font\mybb=msbm10 at 11pt

\def\bb#1{\hbox{\mybb#1}}

\def\bR {\bb{R}}

\def\gX{\Gamma\mkern-4.0mu X}

\def\gY{\Gamma\mkern-2.0mu Y}
\def\gF{\Gamma\mkern-4.0mu F}
\def\gQ{\Gamma\mkern-4.0mu Q}

\def\sX{\slashed {X}}
\def\sgX{\slashed {\gX}}
\def\sY{\slashed {Y}}
\def\sgY{\slashed {\gY}}
\def\sh{\slashed {h}}
\def\sF{\slashed {F}}
\def\sgF{\slashed {\gF}}

\def\sgQ{\slashed {\gQ}}

\newcommand{\bea}{\begin{eqnarray}}
	\newcommand{\eea}{\end{eqnarray}}

\usepackage[usenames,dvipsnames,svgnames,table]{xcolor}	
\usepackage[normalem]{ulem}				
\usepackage{microtype}
\usepackage[colorlinks, citecolor=blue, linkcolor=blue, urlcolor=Maroon, filecolor=Maroon, linktocpage=true]{hyperref} 

\usepackage{chngcntr}
\counterwithout{equation}{section}

\begin{document}

	\begin{center}
		\vspace*{-1.0cm}
		\begin{flushright}
		\end{flushright}

		
		\vspace{2.0cm} {\Large \bf The TCFHs of D=11 AdS backgrounds and hidden symmetries } \\[.2cm]
		
		\vskip 2cm
		G.\,  Papadopoulos and  E.\, P\'erez-Bola\~nos
		\\
		\vskip .6cm


		\begin{small}
			\textit{Department of Mathematics
				\\
				King's College London
				\\
				Strand
				\\
				London WC2R 2LS, UK}
			\\*[.3cm]
			\texttt{george.papadopoulos@kcl.ac.uk}
			\\
			\texttt{edgar.perez$\underline{~}$bolanos@kcl.ac.uk}
		\end{small}
		\\*[.6cm]

	\end{center}

	\vskip 2.5 cm

	\begin{abstract}

We present the TCFHs of all AdS backgrounds of 11-dimensional supergravity and explore some of the properties of the associated connections. Therefore we demonstrate that all Killing spinor bilinears satisfy a generalisation of the conformal Killing-Yano equations with respect to the TCFH connection. In addition we explore the hidden symmetries of spinning particle probes propagating on these backgrounds. We give several examples of hidden symmetries for probes on the maximal supersymmetric AdS backgrounds as well as on some AdS backgrounds that arise as near horizon geometries of intersecting M-branes.
		\noindent
	
	\end{abstract}

	

	\newpage
	
	\renewcommand{\thefootnote}{\arabic{footnote}}

\section{Introduction}

There are two methods to solve the Killing spinor equations (KSEs) of a supergravity theory, the bilinears method \cite{gut1} and the spinorial geometry method \cite{spingeom}; for a review see \cite{rev}. In the latter method the KSEs are solved directly using the (gauge) symmetries of supergravity theories and a realisation of spinors in terms of forms. While in the former method one solves the equations that the KSEs impose on the spacetime form bilinears (Dirac currents) of the Killing spinors. Originally these equations on the form bilinears were not thought to have a geometric interpretation but more recently it was realised that they can be arranged as a twisted covariant form hierarchy (TCFH) \cite{jggp, tcfhgp}. This means that there is a connection ${\cal D}^{\cal F}$ on the space of spacetime forms which depends on the form fluxes ${\cal F}$ of the supergravity theory under consideration such that schematically
\bea
{\cal D}_X^{\cal F}\Omega= i_X {\cal P}+ X\wedge {\cal Q}~,
\label{tcfh}
\eea
for every spacetime vector field\footnote{We have also denoted with $X$ the 1-form dual of $X$ constructed as $X(Y)=g(X,Y)$, where $g$ is the spacetime metric and $Y$ a vector field on the spacetime.} $X$, where $\Omega$ is a multi-form spanned by the form bilinears, and ${\cal P}$ and ${\cal Q}$ are also multi-forms that
depend on the supergravity fields as well as the form bilinears. The connection ${\cal D}^{\cal F}$ is not necessarily form degree preserving. One can also demonstrate that all theories with local supersymmetry, which may include higher order curvature corrections, on spacetimes of any signature  exhibit a TCFH. The presence of a TCFH in a gravity theory with supersymmetry depends  on the structure of the supercovariant connection, and so on the form of gravitino KSE, instead on the field dependence and the couplings of the theory.

A consequence of the TCFH, and equation (\ref{tcfh}) in particular, is that the form bilinears satisfy a generalisation of the conformal Killing-Yano  (CKY) equation\footnote{A $k$-form $\omega$ on a n-dimensional manifold is CKY, iff $\nabla_X\omega=i_X d\omega- (n-k+1)^{-1} X\wedge \delta \omega$, where $\nabla$ is the Levi-Civita connection. In addition, if the form $\omega$  is co-closed, $\delta\omega=0$, then $\omega$ is Killing-Yano. Whereas if $d\omega=0$, then $\omega$ is closed CKY (CCKY).}
with respect to the connection ${\cal D}^{\cal F}$. It is well known for sometime that Killing-Yano (KY) forms  are associated with constants of motion for relativistic particles propagating on a spacetime  \cite{penrose, floyd}. Key applications include the integrability of the geodesic motion as well as that of some classic field equations  in a variety of black holes, see e.g. \cite{carter-b}-\cite{lun} and also reviews \cite{revky, frolov}. For some other applications, see also \cite{kt, gpky, hl, ls1, ls2}.     In addition, KY forms  generate symmetries \cite{gibbons} in spinning particle actions \cite{bvh}.  As a consequence it is natural to ask whether (\ref{tcfh}) gives rise to KY forms which in turn will generate symmetries in various spinning particle probes propagating in supersymmetric backgrounds.  This question has been investigated in several supergravity theories in \cite{ebgp1, ljgp} and several examples have been presented.

The purpose of this paper is to give the TCFHs on the internal space of all warped AdS backgrounds of 11-dimensional supergravity theory.  This will put the conditions on the form bilinears implied by the KSEs in a firm geometric basis. For this we shall use the solution of the (gravitino) KSE of the theory along the AdS subspace of a background presented in \cite{Mads}. Then we shall explore some of the properties of the
TCFH connections which include their (reduced) holonomy for generic internal spaces. In addition we shall examine the conditions under which the form bilinears give rise to
KY or CCKY forms\footnote{The Hodge dual of a CCKY form is a KY form.}. The existence of such forms will imply in turn the presence of constants of motion in the propagation of spinning particles on the internal spaces of such backgrounds. We shall present several backgrounds with KY forms arising from the TCFH of their internal space.  These include the maximally supersymmetric AdS solutions of the theory as well as the near horizon geometries of some intersecting M-brane configurations.

This paper is organised as follows. In sections 2, 3 and 4, we present the TCFH of warped AdS$_2$, AdS$_3$ and AdS$_4$ backgrounds of 11-dimensional supergravity, respectively, and investigate some of the properties of the TCFH connections. In section 5, we give the TCFHs and investigate their properties of the remaining AdS backgrounds. In section 6, we explore the hidden symmetries of probes  that arise from the TCFH of some solutions that include the maximally supersymmetric AdS backgrounds as well as some AdS backgrounds which are the near horizon geometries of intersecting M-branes, and in section 7 we give our conclusions.

\section{The TCFH of warped AdS$_2$ backgrounds}

\subsection{Fields and Killing spinors}

The bosonic fields of 11-dimensional supergravity for warped AdS$_2$ backgrounds, AdS$_2\times_w M^9$, can be written as
\bea
g=2\bbe^+ \bbe^- + \delta_{ij} \bbe^i \bbe^j~,~~~F=\bbe^+ \wedge \bbe^- \wedge Y+X~,
\label{ads2f}
\eea
 where $Y$ and $X$ are a 2-form and 4-form  on the internal space $M^9$ with metric $g(M^9)=\delta_{ij} \bbe^i \bbe^j$, respectively,
\bea
\bbe^+=du ~,~~~ \bbe^-=dr+rh-{1\over2}r^2\ell^{-2} A^{-2} du ~,~~~ {\bf{e}}^i=e^i{}_J dy^J ~,
\label{psfr}
\eea
  is a null pseudo-orthonormal frame on the spacetime with ${\bf{e}}^i$ an orthonormal frame on the internal space $M^9$,  $h=-d\log A^2$ and  $A$ is the warp factor which is a function on $M^9$.  The $y^I$ are coordinates on $M^9$ and $(u,r)$ are the remaining coordinates on the spacetime.  It can be seen after a change of coordinates that the above metric can be rewritten in the standard
warped product  form, $g=A^2 g_\ell(AdS_2)+ g(M^9)$, where $g_\ell(AdS_2)$ is the standard metric on AdS$_2$ of radius $\ell$.

The KSE of 11-dimensional supergravity can be solved along the AdS$_2$ subspace of AdS$_2\times_w M^9$ \cite{Mads} and the Killing spinors $\epsilon$ can expressed as $\epsilon=\epsilon_1+\epsilon_2$ with
\begin{eqnarray}
\epsilon_1 = \phi_- + u \Gamma_+ \Theta_- \phi_- + ru \Gamma_- \Theta_+ \Gamma_+ \Theta_- \phi_-~,~~~
\epsilon_2 = \phi_+ + r \Gamma_- \Theta_+ \phi_+~.
\label{ksads2}
\end{eqnarray}
where $\phi_\pm$ are  spinors that depend only on the coordinates of $M^9$ and satisfy the light-cone
 projections $\Gamma_\pm\phi_\pm=0$, where $\Gamma_\pm$ have been adapted to the frame\footnote{From here on, the gamma matrices are always adapted to a (pseudo-)orthonormal frame.} (\ref{psfr})  and $\Theta_\pm$ are Clifford algebra elements that depend on the fields. For the explicit expressions of $\Theta_\pm$ as well as  for the spinor notation we use below, see \cite{Mads}.  The dependence of Killing spinors in (\ref{ksads2}) on the $(u,r)$ coordinates is explicit as it is that of the fields in (\ref{ads2f}). In addition, the (spacetime) gravitino KSE implies that $\phi_\pm$ satisfy the KSEs
\bea
\label{kse1}
{\cal D}_i^{\pm}\phi_\pm=0~,~~~{\cal D}_i^{\pm}\equiv {\nabla}_i +\Psi_i^{\pm}~,
\eea
on the internal space $M^9$, where ${\nabla}$ is the Levi-Civita connection of the metric $g(M^9)$ induced on the spin bundle and
\bea
\Psi_i^{\pm}=\mp{1\over4}h_i-{1\over 288}\sgX_i+{1\over36}\sX_i\pm{1\over24}\sgY_i\mp{1\over6}\sY_i~.
\eea
 The TCFHs on $M^9$ that we shall explore below are associated with  the supercovariant connections ${\cal D}^{\pm}$.

\subsection{The TCFH on the internal space}

A basis in the space of form bilinears on the internal space $M^9$ is
\bea
\label{ads2 symm}
&&f^{\pm r,s}=\langle \phi_\pm^r, \phi_\pm^s \rangle~,~~~
k^{\pm r,s}=\langle \phi_\pm^r, \Gamma_i\phi_\pm^s\rangle\, \bbe^i~,~~~
\cr
&&
\theta^{\pm r,s}={1\over4!}\langle \phi_\pm^r, \Gamma_{ijk\ell}\phi_\pm^s\rangle\, \bbe^i\wedge \bbe^j\wedge \bbe^k\wedge \bbe^\ell ~,~~~\omega^{\pm r,s}={1\over2}\langle \phi_\pm^r, \Gamma_{ij}\phi_\pm^s\rangle\, \bbe^i\wedge \bbe^j~,~~~
\cr
&&\varphi^{\pm r,s}={1\over 3!}\langle \phi_\pm^r, \Gamma_{ijk}\phi_\pm^s\rangle\, \bbe^i\wedge \bbe^j \wedge \bbe^k~,
\eea
where all (time-like) space-like gamma matrices are (anti-)Hermitian with respect to the inner product $\langle \cdot, \cdot\rangle$.
As the Killing spinors of 11-dimensional supergravity are real note that the bilinears $f$, $k$ and $\theta$ are symmetric in the exchange of spinors $\phi_\pm^r$ and $\phi_\pm^s$ and the rest are skew-symmetric.

The TCFH  associated with the supercovariant connections ${\cal D}^\pm$ on the bilinears that are symmetric in the exchange of $\phi_\pm^r$ with $\phi_\pm^s$ is
\bea
&&
{\cal D}^{(\pm)\cal F}_i f^{\pm} \equiv \nabla_i f^\pm \pm\partial_i \log A \, f^\pm=\pm{1\over144}\,{}^*X_{i\ell_1\ell_2\ell_3\ell_4}\theta^\pm{}^{\ell_1\ell_2\ell_3\ell_4}\pm{1\over3}Y_{i\ell}k^\pm{}^{\ell}~,~~~
\cr
&&
{\cal D}^{(\pm)\cal F}_i k_j^{\pm}\equiv \nabla_i k^\pm_j\pm\partial_i\log A \, k^\pm_j-{1\over12}X_{i\ell_1\ell_2\ell_3}\theta^\pm{}^{\ell_1\ell_2\ell_3}{}_{j}
\cr
&&
={1\over18}X_{\ell_1\ell_2\ell_3[i}\theta^\pm{}^{\ell_1\ell_2\ell_3}{}_{j]}+{1\over144}\delta_{ij}X_{\ell_1\ell_2\ell_3\ell_4}\theta^\pm{}^{\ell_1\ell_2\ell_3\ell_4}
\pm{1\over3}Y_{ij}f^\pm\pm{1\over12}Y_{\ell_1\ell_2}\theta^\pm{}^{\ell_1\ell_2}{}_{ij}~,~~~
\cr
&&
{\cal D}^{(\pm)\cal F}_i \theta_{j_1j_2j_3j_4}^{\pm}\equiv \nabla_i \theta^\pm_{j_1j_2j_3j_4}\pm\partial_i\log A \, \theta^\pm_{j_1j_2j_3j_4}\pm{3\over2}\,{}^*X_{\ell_1\ell_2i[j_1j_2}\theta^\pm{}^{\ell_1\ell_2}{}_{j_3j_4]}
\cr
&&
+2X_{i[j_1j_2j_3}k^\pm_{j_4]}+{1\over3}\,{}^*Y_{\ell_1\ell_2\ell_3i[j_1j_2j_3}\theta^\pm{}^{\ell_1\ell_2\ell_3}{}_{j_4]}
\cr
&&
=\pm{5\over3}\,{}^*X_{\ell_1\ell_2[ij_1j_2}\theta^\pm{}^{\ell_1\ell_2}{}_{j_3j_4]}+{5\over6}X_{[ij_1j_2j_3}k^\pm_{j_4]}\pm{1\over6}\,{}^*X_{ij_1j_2j_3j_4}f^\pm
\cr
&&
\mp{2\over3}\delta_{i[j_1}\,{}^*X_{j_2j_3|k_1k_2k_3}\theta^\pm{}^{\ell_1\ell_2\ell_3}{}_{j_4]}+{2\over3}\delta_{i[j_1}X_{j_2j_3j_4]\ell}k^\pm{}^{\ell}
-{1\over18}\delta_{i[j_1}\,{}^*Y_{j_2j_3j_4]\ell_1\ell_2\ell_3\ell_4}\theta^\pm{}^{\ell_1\ell_2\ell_3\ell_4}
\cr
&&
+{5\over18}\,{}^*Y_{[ij_1j_2j_3|\ell_1\ell_2\ell_3|}\theta^\pm{}^{\ell_1\ell_2\ell_3}{}_{j_4]}\mp 2\delta_{i[j_1}Y_{j_2j_3}k^\pm_{j_4]}~.
\label{tcfhads21}
\eea
While the TCFH on the bilinears which are skew-symmetric in the exchange of $\phi_\pm^r$ with $\phi_\pm^s$ is
\bea
&&
{\cal D}^{(\pm)\cal F}_i\omega_{j_1j_2}^{\pm}\equiv \nabla_i \omega^\pm_{j_1j_2}\pm\partial_i \log A \, \omega^\pm_{j_1j_2}-{1\over2}X_{\ell_1\ell_2i[j_1}\varphi^\pm{}^{\ell_1\ell_2}{}_{j_2]}\mp{1\over2}Y_{i\ell}\varphi^\pm{}^{\ell}{}_{j_1j_2}
\cr
&&
=-{1\over4}X_{\ell_1\ell_2[ij_1}\varphi^\pm{}^{\ell_1\ell_2}{}_{j_2]}\mp{1\over12}\,{}^*X_{ij_1j_2\ell_1\ell_2}\omega^\pm{}^{\ell_1\ell_2}
-{1\over18}\delta_{i[j_1}X_{j_2]\ell_1\ell_2\ell_3}\varphi^\pm{}^{\ell_1\ell_2\ell_3}
\cr
&&
\pm{1\over6}\delta_{i[j_1}\varphi^\pm{}^{\ell_1\ell_2}{}_{j_2]}Y_{\ell_1\ell_2}\pm{1\over2}Y_{\ell[i}\varphi^\pm{}^{\ell}{}_{j_1j_2]}~,~~~
\cr
&&
{\cal D}^{(\pm)\cal F}_i\varphi_{j_1j_2j_3}^{\pm}\equiv \nabla_i \varphi^\pm_{j_1j_2j_3}\pm\partial_i\log A\,\varphi^\pm_{j_1j_2j_3}\pm{3\over4}\,{}^*X_{\ell_1\ell_2i[j_1j_2}\varphi^\pm{}^{\ell_1\ell_2}{}_{j_3]}
\cr
&&
-{3\over2}X_{\ell i[j_1j_2}\omega^\pm{}^{\ell}{}_{j_3]}\mp{3\over2}Y_{i[j_1}\omega^\pm_{j_2j_3]}
\cr
&&
=\pm{2\over3}\,{}^*X_{\ell_1\ell_2[ij_1j_2}\varphi^\pm{}^{\ell_1\ell_2}{}_{j_3]}-{2\over3}X_{\ell[ij_1j_2}\omega^\pm{}^{\ell}{}_{j_3]}
\pm{1\over6}\delta_{i[j_1}\,{}^*X_{j_2j_3]\ell_1\ell_2\ell_3}\varphi^\pm{}^{\ell_1\ell_2\ell_3}
\cr
&&
-{1\over4}\delta_{i[j_1}X_{j_2j_3]\ell_1\ell_2}\omega^\pm{}^{\ell_1\ell_2}-{1\over36}\,{}^*Y_{ij_1j_2j_3\ell_1\ell_2\ell_3}\varphi^\pm{}^{\ell_1\ell_2\ell_3}\mp Y_{[ij_1}\omega^\pm_{j_2j_3]}\mp\delta_{i[j_1}Y_{j_2|\ell|}\omega^\pm{}^{\ell}{}_{j_3]}~,
\label{tcfhads22}
\eea
where  ${}^* X$ is the Hodge dual\footnote{In our conventions $\Gamma_{i_1\dots i_9}\phi_{\pm}=\pm\epsilon_{i_1\dots i_9}\phi_{\pm}$ with  $\epsilon^{123456789}=1$.} of $X$ and similarly for the other fields. In the TCFH above we have  suppressed the $r,s$ indices on the bilinears that count the number of Killing spinors.

To compute (\ref{tcfhads21}) and (\ref{tcfhads22}), one utilises the identity 
\bea
&&\nabla_i\langle \phi^r_\pm, \Gamma_{i_1\dots i_k} \phi^s_\pm\rangle=\langle \nabla_i\phi^r_\pm, \Gamma_{i_1\dots i_k} \phi^s_\pm\rangle+
\langle \phi^r_\pm, \Gamma_{i_1\dots i_k} \nabla_i\phi^s_\pm\rangle
\cr
&&\qquad \qquad
=-\langle \Psi_i^{\pm}\phi^r_\pm, \Gamma_{i_1\dots i_k} \phi^s_\pm\rangle-\langle \phi^r_\pm, \Gamma_{i_1\dots i_k} \Psi_i^{\pm}\phi^s_\pm\rangle~,
\eea
where  the last equality follows after a use the KSEs (\ref{kse1}). Then one proceed with some extensive Clifford algebra computation.

The TCFH above has been expressed in terms of the minimal connection ${\cal D}^{(\pm)\cal F}$, see \cite{tcfhgp} for the definition. As expected  ${\cal D}^{(\pm)\cal F}$ is form degree non-preserving connection. As the action of ${\cal D}^{(\pm)\cal F}_i$ on the space of forms preserves the subspaces of symmetric and skew-symmetric bilinears in the exchange of $\phi^r$ and $\phi^s$ Killing spinors and acts trivially on the scalars $f$, the reduced holonomy\footnote{From here on with the term holonomy we shall always refer to the reduced holonomy of the TCFH connection, i.e. the connected to the identity part of the holonomy group.} of the connection is included in (the connected to identity component of)  $GL(135)\times GL(120)$. Note that the  holonomy of the maximal TCFH connection, see again \cite{tcfhgp},  is contained in $GL(136)\times GL(120)$ as it acts non-trivially on the scalars.

An alternative way to see that the holonomy of ${\cal D}^{(\pm)\cal F}$ is included in $GL(135)\times GL(120)$ is to observe that $\phi_\pm$ can be thought of as Majorana spinors  of $\mathfrak{spin}(9)$. It is well known that the tensor product of two $\mathfrak{spin}(9)$ Majorana representations, $\Delta_{\bf 16}$, decomposes as
\bea
\Delta_{\bf 16}\otimes \Delta_{\bf 16}=\oplus^4_{k=0} \Lambda^k (\bR^9)~,
\eea
where $\Lambda^k (\bR^9)$ is the irreducible representation of $\mathfrak{spin}(9)$ on the space of $k$-degree forms on $\bR^9$.
The action of the supercovariant connection on the tensor product of two spin bundles, i.e. on a bispinor, preserves the symmetric and skew symmetric subspaces.  As the rank of the spin bundle is 16, these sub-bundles have rank 136 and 120, respectively. So the holonomy of all connections of the TCFH is included in $GL(136)\times GL(120)$.

Although the form bilinears are CKY forms with respect to the
TCFH connection ${\cal D}^{(\pm)\cal F}$  as expected, it is clear from the TCFH (\ref{tcfhads21}) and (\ref{tcfhads22}) that they   are neither KY nor CCKY forms for generic supersymmetric backgrounds. However we shall demonstrate that for special solutions several terms in the TCFH  vanish and as a result some bilinears become either KY or  CCKY forms.

\section{The TCFH of warped AdS$_3$ backgrounds}

\subsection{Fields and Killing spinors}

 The bosonic fields of  11-dimensional supergravity for a warped AdS$_3$ background, AdS$_3\times_w M^8$, can be written as
\bea
g = 2 \bbe^+ \bbe^- + (\bbe^z)^2  + \delta_{ij} \bbe^i \bbe^j~,~~~
  F= \bbe^+\wedge \bbe^-\wedge dz\wedge Q+X~,
 \label{ads3f}
\eea
where the metric, $g(M^8)$, on the internal space $M^8$ is $g(M^8)= \delta_{ij} \bbe^i \bbe^j$ and $Q$ and $X$ are a 1-form and a 4-form on  $M^8$, respectively. Moreover
\bea
\bbe^+=du~,~~~\bbe^-= dr - 2 \ell^{-1} r dz - 2 r d \ln A~,~~~\bbe^z= A dz~,~~~\bbe^i=e^i_J dy^J~,
\eea
is a  null pseudo-orthonormal frame on the spacetime  with $\bbe^i$ an orthonormal frame on $M^8$, $y$ are  coordinates on $M^8$ and $(u,r,z)$  are  the rest of coordinates of spacetime  and $A$ is the warp factor. As for the AdS$_2$ backgrounds in the previous section,  there is a coordinate transformation such that the spacetime metric $g$ can be put into the standard warped form, $g=A^2 g_\ell(AdS_3)+ g(M^8)$, where $g_\ell(AdS_3)$ is the standard metric on AdS$_3$ with radius $\ell$.

The gravitino KSE of 11-dimensional supergravity can be solved \cite{Mads} along the AdS$_3$ subspace of a  AdS$_3\times_w M^8$   background with fields (\ref{ads3f}).  The Killing spinors $\epsilon$  can be expressed as,
$
\epsilon=\epsilon(\sigma_\pm)+\epsilon(\tau_\pm)
$,
with
\bea
\epsilon(\sigma_\pm)&=&\sigma_++\sigma_--\ell^{-1} A^{-1} u \Gamma_{+z} \sigma_-~,
\cr
\epsilon(\tau_\pm)&=&e^{-{z\over\ell}}\tau_+-\ell^{-1} A^{-1} r e^{-{z\over\ell}} \Gamma_{-z} \tau_++e^{{z\over \ell}} \tau_-~,
\label{ksp2}
\eea
 where $\sigma_\pm$ and $\tau_\pm$  spinors satisfy the lightcone projections $\Gamma_\pm\sigma_\pm=\Gamma_\pm\tau_\pm=0$, and depend only on the coordinates of $M^8$. In addition,  they satisfy the KSEs
\bea
{\cal D}^{\pm}_i\sigma_\pm={\cal D}^{\pm}_i \tau_\pm=0~,
\eea
on the internal space $M^8$, where
\bea
{\cal D}^{\pm}_i \equiv \nabla_i \pm {1\over 2}\partial_i \log A-{1\over288}\sgX_i+{1\over36}\sX_i \mp {1\over12}A^{-1}\Gamma_z\sgQ_i\pm {1\over6} A^{-1}\Gamma_zQ_i~,
\eea
are the supercovariant connections on $M^8$ and $\nabla$ is the connection induced on the spin bundle from the Levi-Civita connection of the metric $g(M^8)$.  Furthermore $\sigma_\pm$ and $\tau_\pm$ satisfy an additional algebraic KSE on $M^8$ arising from the integration of the
gravitino KSE of 11-dimensional supergravity along the $z$ coordinate. These algebraic KSEs have been explained in detail  in \cite{Mads} and they will be used in the examples below to produce the right counting for the number of Killing spinors of the AdS backgrounds but they do not contribute in the TCFH below.

\subsection{The TCFH on the internal space}

A basis in the space of form bilinear  on the internal space $M^8$ is
\bea
\label{ads3 symm}
&&
f^{\pm r,s}=\langle \phi_\pm^r, \phi_\pm^s \rangle~,~~~\tilde{f}^{\pm r,s}=\langle \phi_\pm^r,\Gamma_{z} \phi_\pm^s \rangle~,~~~
k^{\pm r,s}=\langle \phi_\pm^r, \Gamma_i\phi_\pm^s\rangle\, \bbe^i~,~~~
\cr
&&
\theta^{\pm r,s}={1\over 4!}\langle \phi_\pm^r, \Gamma_{ijk\ell}\phi_\pm^s\rangle\, \bbe^i \wedge \bbe^j \wedge \bbe^k \wedge \bbe^\ell ~,~~
\tilde{\varphi}^{\pm r,s}={1\over 3!}\langle \phi_\pm^r, \Gamma_{ijk}\Gamma_{z}\phi_\pm^s\rangle\, \bbe^i \wedge \bbe^j \wedge \bbe^k~,~~~
\cr
&&
\tilde{k}^{\pm r,s}=\langle \phi_\pm^r, \Gamma_i\Gamma_{z}\phi_\pm^s\rangle\, \bbe^i~,~~~
\omega^{\pm r,s}={1\over2}\langle \phi_\pm^r, \Gamma_{ij}\phi_\pm^s\rangle\, \bbe^i \wedge \bbe^j~,~~~
\cr
&& \tilde{\omega}^{\pm r,s}={1\over2}\langle \phi_\pm^r, \Gamma_{ij}\Gamma_{z}\phi_\pm^s\rangle\, \bbe^i \wedge \bbe^j~,~~~
\varphi^{\pm r,s}={1\over 3!}\langle \phi_\pm^r, \Gamma_{ijk}\phi_\pm^s\rangle\, \bbe^i \wedge \bbe^j \wedge \bbe^k ~,~~~
\eea
where $\phi_\pm$ stands\footnote{One can also consider mixed $\sigma_\pm$ and $\tau_\pm$ form bilinears.  The TCFH is the same as the one stated below for the form bilinear basis above.} for either $\sigma_\pm$ or $\tau_\pm$. The first five form bilinears are symmetric in the exchange of $\phi^r_\pm$ with $\phi^s_\pm$ while the rest are skew-symmetric.

The TCFH expressed\footnote{In our conventions $\Gamma_{z}\Gamma_{i_1\dots i_8}\phi_{\pm}=\pm\epsilon_{i_1\dots i_8}\phi_{\pm}$ with $\epsilon^{12345678}=1$.}  in terms of the minimal connection is
\bea
&&
{\cal D}^{(\pm)\cal F}_i f^{\pm}\equiv \nabla_i f^\pm \pm\partial_i \log A \, f^\pm=\pm{1\over36}\,{}^*X_{i}{}^{\ell_1\ell_2\ell_3}\tilde{\varphi}^\pm_{\ell_1\ell_2\ell_3}\mp{1\over3}A^{-1}Q_i\tilde{f}^\pm~,
\cr
&&
{\cal D}^{(\pm)\cal F}_i k_j^{\pm}\equiv \nabla_i k^\pm_j\pm\partial_i\log A \, k^\pm_j-{1\over12}X_{i\ell_1\ell_2\ell_3}\theta^\pm{}^{\ell_1\ell_2\ell_3}{}_{j}
\cr
&&
=-{1\over18}X_{[i|\ell_1\ell_2\ell_3|}\theta^\pm{}^{\ell_1\ell_2\ell_3}{}_{j]}
+{1\over144}\delta_{ij}X_{\ell_1\ell_2\ell_3\ell_4}\theta^\pm{}^{\ell_1\ell_2\ell_3\ell_4}\mp{1\over6}A^{-1}Q_{\ell}\tilde{\varphi}^\pm{}^{\ell}{}_{ij}~,~~~
\cr
&&
{\cal D}^{(\pm)\cal F}_i\theta_{j_1j_2j_3j_4}^{\pm}\equiv \nabla_i \theta^\pm_{j_1j_2j_3j_4}\pm\partial_i \log A\mp3\,{}^*X_{\ell i[j_1j_2}\tilde{\varphi}^\pm{}^{\ell}{}_{j_3j_4]}+2X_{i[j_1j_2j_3}k^\pm_{j_4]}
\cr
&&
+{1\over3}A^{-1}\,{}^*Q_{\ell_1\ell_2\ell_3i[j_1j_2j_3}\theta^\pm{}^{\ell_1\ell_2\ell_3}{}_{j_4]}
\cr
&&
=\mp{10\over3}\,{}^*X_{\ell[ij_1j_2}\tilde{\varphi}^\pm{}^{\ell}{}_{j_3j_4]}+{5\over6}X_{[ij_1j_2j_3}k^\pm_{j_4]}\mp2\delta_{i[j_1}\,{}^*X_{j_2j_3|\ell_1\ell_2|}
\tilde{\varphi}^\pm{}^{\ell_1\ell_2}{}_{j_4]}+{2\over3}\delta_{i[j_1}X_{j_2j_3j_4]\ell}k^\pm{}^{\ell}
\cr
&&
-{1\over18}A^{-1}\delta_{i[j_1}\,{}^*Q_{j_2j_3j_4]\ell_1\ell_2\ell_3\ell_4}\theta^\pm{}^{\ell_1\ell_2\ell_3\ell_4}
+{5\over18}A^{-1}\,{}^*Q_{\ell_1\ell_2\ell_3[ij_1j_2j_3}\theta^\pm{}^{\ell_1\ell_2\ell_3}{}_{j_4]}~,~~~
\cr
&&
{\cal D}^{(\pm)\cal F}_i\tilde{\varphi}_{j_1j_2j_3}^{\pm}\equiv \nabla_i \tilde{\varphi}^\pm_{j_1j_2j_3}\pm\partial_i \log A \, \tilde{\varphi}^\pm_{j_1j_2j_3}\pm{3\over4}\,{}^*X_{\ell_1\ell_2i[j_1}\theta^\pm{}^{\ell_1\ell_2}{}_{j_2j_3]}
\cr
&&
=\pm\,{}^*X_{\ell_1\ell_2[ij_1}\theta^\pm{}^{\ell_1\ell_2}{}_{j_2j_3]}-{1\over3}X_{ij_1j_2j_3}\tilde{f}^\pm \pm{1\over6}\,{}^*X_{ij_1j_2j_3}f^\pm \pm{1\over3}\delta_{i[j_1}\,{}^*X_{j2|\ell_1\ell_2\ell_3|}\theta^\pm{}^{\ell_1\ell_2\ell_3}{}_{j_3]}
\cr
&&
-{1\over36}A^{-1}\,{}^*Q_{ij_1j_2j_3\ell_1\ell_2\ell_3}\tilde{\varphi}^\pm{}^{\ell_1\ell_2\ell_3}\mp A^{-1}\delta_{i[j_1}Q_{j_2}k^\pm_{j_3]}~,
\cr
&&
{\cal D}^{(\pm)\cal F}_i \tilde{f}^{\pm}\equiv \nabla_i \tilde{f}^\pm \pm\partial_i \log A \tilde{f}^\pm ={1\over18}X_{i\ell_1\ell_2\ell_3}\tilde{\varphi}^\pm{}^{\ell_1\ell_2\ell_3}\mp{1\over3}A^{-1}Q_i f^\pm~,
\cr
&&
{\cal D}^{(\pm)\cal F}_i\omega_{j_1j_2}^{\pm}\equiv \nabla_i \omega^\pm_{j_1j_2}\pm\partial_i \log A \omega^\pm_{j_1j_2}-{1\over2}X_{\ell_1\ell_2i[j_1}\varphi^\pm{}^{\ell_1\ell_2}{}_{j_2]}\pm{1\over2}A^{-1}Q_i\tilde{\omega}^\pm_{j_1j_2}
\cr
&&
=-{1\over4}X_{\ell_1\ell_2[ij_1}\varphi^\pm{}^{\ell_1\ell_2}{}_{j_2]}-{1\over18}\delta_{i[j_1}X_{j_2]\ell_1\ell_2\ell_3}\varphi^\pm{}^{\ell_1\ell_2\ell_3}
\mp{1\over6}\,{}^*X_{ij_1j_2\ell}\tilde{k}^\pm{}^{\ell}
\cr
&&
\pm{1\over3}A^{-1}\delta_{i[j_1}\tilde{\omega}^\pm{}^{\ell}{}_{j_2]}Q_{\ell}
\pm{1\over2}A^{-1}Q_{[i}\tilde{\omega}^\pm_{j_1j_2]}~,~~~
\cr
&&
{\cal D}^{(\pm)\cal F}_i\varphi_{j_1j_2j_3}^{\pm}\equiv \nabla_i \varphi^\pm_{j_1j_2j_3}\pm\partial_i \log A \varphi^\pm_{j_1j_2j_3}-{3\over2}X_{\ell i[j_1j_2}\omega^\pm{}^{\ell}{}_{j_3]}\pm{3\over2}\,{}^*X_{\ell i[j_1j_2}\tilde{\omega}^\pm{}^{\ell}{}_{j_3]}
\cr
&&
=\pm{4\over3}\,{}^*X_{\ell[ij_1j_2}\tilde{\omega}^\pm{}^{\ell}{}_{j_3]}-{2\over3}X_{\ell[ij_1j_2}\omega^\pm{}^{\ell}{}_{j_3]}
\pm{1\over2}\delta_{i[j_1}\,{}^*X_{j_2j_3]\ell_1\ell_2}\tilde{\omega}^\pm{}^{\ell_1\ell_2}-{1\over4}\delta_{i[j_1}X_{j_2j_3]\ell_1\ell_2}\omega^\pm{}^{\ell_1\ell_2}
\cr
&&
-{1\over36}A^{-1}\,{}^*Q_{ij_1j_2j_3\ell_1\ell_2\ell_3}\varphi^\pm{}^{\ell_1\ell_2\ell_3}\mp A^{-1}\delta_{i[j_1}Q_{j_2}\tilde{k}^\pm_{j_3]}~,~~~
\cr
&&
{\cal D}^{(\pm)\cal F}_i \tilde{k}_j^{\pm}\equiv \nabla_i \tilde{k}^\pm_j\pm\partial_i\log A \tilde{k}^\pm_j={1\over6}X_{ij\ell_1\ell_2}\tilde{\omega}^\pm{}^{\ell_1\ell_2}\mp{1\over12}\,{}^*X_{ij\ell_1\ell_2}\omega^\pm{}^{\ell_1\ell_2}\pm {1\over6}A^{-1}Q_{\ell}\varphi^\pm{}^{\ell}{}_{ij}~,~~~
\cr
&&
{\cal D}^{(\pm)\cal F}_i\tilde{\omega}_{j_1j_2}^{\pm}\equiv \nabla_i \tilde{\omega}^\pm_{j_1j_2}\pm\partial_i \log A \tilde{\omega}^\pm_{j_1j_2}\mp{1\over2}\,{}^*X_{\ell_1\ell_2i[j_1}\varphi^\pm{}^{\ell_1\ell_2}{}_{j_2]}\pm{1\over2}A^{-1}Q_i\omega^\pm_{j_1j_2}
\cr
&&
=\mp{1\over2}\,{}^*X_{\ell_1\ell_2[ij_1}\varphi^\pm{}^{\ell_1\ell_2}{}_{j_2]}-{1\over3}X_{ij_1j_2\ell}\tilde{k}^\pm{}^{\ell}
\mp{1\over9}\delta_{i[j_1}\,{}^*X_{j_2]\ell_1\ell_2\ell_3}\varphi^\pm{}^{\ell_1\ell_2\ell_3}\pm{1\over2}A^{-1}Q_{[i}\omega^\pm_{j_1j_2]}
\cr
&&
\pm{1\over3}A^{-1}\delta_{i[j_1}\omega^\pm{}^{\ell}{}_{j_2]}Q_{\ell}~,
\eea
where ${}^*X$ is the Hodge dual of $X$ on the internal space $M^8$ and similarly for the other fields. We have also suppressed the $r,s$ indices on the form bilinears that count the number of Killing spinors.

The action of the minimal TCFH connection on the space of forms preserves the subspaces of forms with are symmetric and skew-symmetric in the exchange of the Killing spinors $\phi^r_\pm$ and $\phi^s_\pm$. Furthermore, it preserves the subspaces of 1-forms  $\tilde k^\pm$, where it acts as the Levi-Civita connection up to a rescalling with the warp factor $A$,  and acts trivially on the scalars $f^\pm$ and $\tilde f^\pm$. Therefore the reduced holonomy is included in  $GL(134)\times SO(8)\times GL(112)$.  The reduced holonomy of the maximal TCFH connection instead is included in $GL(136)\times GL(120)$ because it preserves only the subspaces
of forms which are symmetric and skew-symmetric in the exchange of the Killing spinors $\phi^r_\pm$ and $\phi^s_\pm$.

The holonomy of the TCFH connection can be understood in a way similar to that of AdS$_2$ backgrounds. As $\phi_\pm$ can be viewed as Majorana $\mathfrak{spin}(8)$ spinors, it is known that the product of two such Majorana representations, $\Delta_{\bf 16}$,
can be decomposed in terms of form representations, $\Lambda^k(\bR^8)$, of $\mathfrak{spin}(8)$ as
\bea
\Delta_{\bf 16} \otimes \Delta_{\bf 16}=\oplus_{k=0}^8 \Lambda^k(\bR^8)~.
\eea
As the supercovariant derivative preserves the space of symmetric and skew-symmetric bi-spinors, it is clear that the  holonomy  of all TCFH connections
will be included in $GL(136)\times GL(120)$, where $136$ is the rank of the sub-bundle of symmetric bi-spinors while $120$ is the rank of the sub-bundle of skew-symmetric bi-spinors.

As expected all form bilinears are CKY forms with respect to the TCFH connections ${\cal D}^{(\pm)\cal F}$ in agreement with the general result in \cite{tcfhgp}. Apart from $A^{\pm1} \tilde k^\pm$ which is a Killing 1-form, the TCFH does not imply  that the remaining form bilinears are KY forms for generic supersymmetric backgrounds.  However, we shall demonstrate that many of them are either KY or CCKY forms for some AdS$_3$ solutions of 11-dimensional supergravity.

\section{The TCFH of warped AdS$_4$ backgrounds}

\subsection{Fields and Killing spinors}

The bosonic fields of 11-dimensional supergravity of warped  AdS$_4$ backgrounds, $AdS_4 \times_w M^7$, can be written as
\bea
&&
g=2\bbe^+ \bbe^- + ({\bf{e}}^z)^2+({\bf{e}}^x)^2+ g(M^7)~,~~~
F=S\, {\bf{e^+ \wedge e^-}} \wedge \bbe^z\wedge \bbe^x+ X~,
\label{fads4}
\eea
with $g(M^7)= \delta_{ij} {\bf{e}}^i{\bf{e}}^j$,  where S and X are a 0-form and 4-form on the internal space $M^7$, respectively. In addition,
\bea
&&
{\bf{e^+}}=du ~,~~~ {\bf{e^-}}=dr-2\ell^{-1}rdz-2rd\ln A ~,~~~ {\bf{e}}^z=Adz ~,~~~ {\bf{e}}^x=A e^{z\over \ell}dx ~,~~~
\cr
&&
{\bf{e}}^i=e^i{}_J dy^J ~,
\eea
is a null pseudo-orthonormal frame on the spacetime with $\bbe^i$ an orthonormal frame on $M^7$,  $y$ are coordinates on $M^7$ and $(u, r, z, x)$  are the remaining  coordinates of the spacetime, and $\ell$ is the radius of AdS$_4$. As in previous cases, after a coordinate transformation the spacetime metric can be written in the standard warped form with warp factor $A^2$.

The gravitino KSE of 11-dimensional supergravity  can be explicitly integrated along the $(u,r,z,x)$ coordinates and the Killing spinors can be written as, $\epsilon=\epsilon(\sigma_\pm)+\epsilon(\tau_\pm)$, see \cite{Mads},
with
\bea
&&
\epsilon(\sigma_\pm)=\sigma_+ +\sigma_- -\ell^{-1}e^{z\over\ell}x\Gamma_{xz}\sigma_- -\ell^{-1}A^{-1}u\Gamma_{+z}\sigma_-~,~~~
\cr
&&
\epsilon(\tau_\pm)=e^{-{z\over\ell}}\tau_+  -\ell^{-1}A^{-1}re^{z\over\ell}\Gamma_{-z}\tau_+-\ell^{-1}x\Gamma_{xz}\tau_+ +e^{z\over\ell}\tau_-~,
\eea
 where the spinors $\sigma_\pm$ and $\tau_\pm$ depend only on the coordinates of $M^7$ and satisfy the light-cone projections $\Gamma_\pm \sigma_\pm=0$ and $\Gamma_\pm \tau_\pm=0$. Furthermore, these spinors satisfy the KSEs
\bea
{\cal D}_i^{\pm}\sigma_\pm ~,~~~ {\cal D}_i^\pm\tau_\pm=0~,
\eea
on the internal space $M^7$, where the supercovariant connection is
\bea
{\cal D}^{\pm}_i \equiv \nabla_i \pm {1\over 2}\partial_i \log A-{1\over288}\sgX_i+{1\over36}\sX_i \pm {1\over12}S\Gamma_{izx}~,
\eea
with $\nabla$  induced by the Levi-Civita connection of the metric $g(M^7)$ on $M^7$. The spinors $\sigma_\pm$ and $\tau_\pm$ satisfy an additional algebraic KSE which arises from the integration of the gravitino KSE of 11-dimensional supergravity along the $z$ coordinate. These algebraic KSEs can be found in \cite{Mads} and they are essential for the correct counting of Killing spinors for warped AdS backgrounds.

\subsection{The TCFH on the internal manifold}

A basis in the space of form bilinears on the internal space $M^7$ is
\bea
&&
f^\pm{}^{r,s}=\langle \phi_\pm^r, \phi_\pm^s \rangle~,~~~
k^{\pm r,s}=\langle \phi_\pm^r, \Gamma_i\phi_\pm^s\rangle\, {\bf{e}}^i~,~~~
\tilde{\omega}^{\pm r,s}={1\over2}\langle \phi_\pm^r, \Gamma_{ij}\Gamma_{zx}\phi_\pm^s\rangle\, {\bf{e}}^i \wedge {\bf{e}}^j ~,~~~
\cr
&&
\tilde{\varphi}^{\pm r,s}={1\over3!}\langle \phi_\pm^r, \Gamma_{ijk}\Gamma_{zx}\phi_\pm^s\rangle\, {\bf{e}}^i\wedge {\bf{e}}^j \wedge {\bf{e}}^k~,~~~ \tilde{k}^{\pm r,s}=\langle \phi_\pm^r, \Gamma_i\Gamma_{zx}\phi_\pm^s\rangle\, {\bf{e}}^i~,~~~
\cr
&&
\omega^{\pm r,s}={1\over2}\langle \phi_\pm^r, \Gamma_{ij}\phi_\pm^s\rangle\, {\bf{e}}^i \wedge {\bf{e}}^j~,~~~
\varphi^{\pm r,s}={1\over3!}\langle \phi_\pm^r, \Gamma_{ijk}\phi_\pm^s\rangle\, {\bf{e}}^i\wedge {\bf{e}}^j \wedge {\bf{e}}^k ~,~~~
\cr &&
\tilde{f}^{\pm r,s}=\langle \phi_\pm^r,\Gamma_{zx} \phi_\pm^s \rangle ~.~~~
\eea
where $\phi_\pm$ stands\footnote{Unlike the case of warped AdS$_3$ backgrounds, the $\sigma_\pm$ and $\tau_\pm$ Killing spinors of all warped AdS$_k$, $k>3$, backgrounds are related with  Clifford algebra operations.} for either $\sigma_\pm$ or $\tau_\pm$.
Note that the bilinears $f$, $k$ $\tilde{\omega}$ and $\tilde{\varphi}$ are symmetric in the exchange of spinors $\phi_\pm^r$ and $\phi_\pm^s$ while the rest are skew-symmetric.

The TCFH expressed in terms of the minimal connection can be written as\footnote{In our conventions $\Gamma_{zx}\Gamma_{i_1i_2i_3i_4i_5i_6i_7}\phi_{\pm}=\pm\epsilon_{i_1i_2i_3i_4i_5i_6i_7}\phi_{\pm}$ with $\epsilon^{1234567}=1$}
\bea
&&
{\cal D}^{(\pm){\cal F}}_i f^{\pm}\equiv \nabla_i f^\pm\pm\partial_i \log A f^\pm=\pm{1\over12}\,{}^*X_{i\ell_1\ell_2}\tilde{\omega}^{\pm \ell_1\ell_2}~,
\cr
&&
{\cal D}^{(\pm){\cal F}}_i k_j^{\pm}\equiv \nabla_i k^\pm_j\pm\partial_i\log A k^\pm_j\mp{1\over4}\,{}^*X_{i\ell_1\ell_2}\tilde{\varphi}^{\pm\ell_1\ell_2}{}_{j}
\cr
&&
=\pm{1\over6}S\tilde{\omega}^\pm_{ij}\mp{1\over3}{}^*X_{\ell_1\ell_2[i}\tilde{\varphi}^{\pm \ell_1\ell_2}{}_{j]}\mp{1\over18}\delta_{ij}\,{}^*X_{\ell_1\ell_2\ell_3}\tilde{\varphi}^{\pm\ell_1\ell_2\ell_3}~,~~~
\cr
&&
{\cal D}^{(\pm){\cal F}}_i\tilde{\omega}_{j_1j_2}^{\pm}\equiv \nabla_i\tilde{\omega}^\pm_{j_1j_2}\pm\partial_i \log A \tilde{\omega}^\pm_{j_1j_2}-{1\over2}X_{\ell_1\ell_2i[j_1}\tilde{\varphi}^{\pm\ell_1\ell_2}{}_{j_2]}
\cr
&&
=-{1\over4}X_{\ell_1\ell_2[ij_1}\tilde{\varphi}^{\pm\ell_1\ell_2}{}_{j_2]}
-{1\over18}\delta_{i[j_1}X_{j_2]\ell_1\ell_2\ell_3}\tilde{\varphi}^{\pm \ell_1\ell_2\ell_3}\pm{1\over6}\,{}^*X_{ij_1j_2}f^\pm\mp{1\over3}S\delta_{i[j_1}k^\pm_{j_2]}~,~~~
\cr
&&
{\cal D}^{(\pm){\cal F}}_i \tilde{\varphi}_{j_1j_2j_3}^{\pm}\equiv \nabla_i \tilde{\varphi}^\pm_{j_1j_2j_3}\pm\partial_i \log A \tilde{\varphi}^\pm_{j_1j_2j_3}-{3\over2}X_{\ell i[j_1j_2}\bar{\omega}^{\pm\ell}{}_{j_3]}\mp{3\over2}\,{}^*X_{i[j_1j_2}k^\pm_{j_3]}
\cr
&&
=\mp{4\over3}\,{}^*X_{[ij_1j_2}k^\pm_{j_3]}-{2\over3}X_{\ell[ij_1j_2}\tilde{\omega}^{\pm\ell}{}_{j_3]}\mp\delta_{i[j_1}\,{}^*X_{j_2j_3]\ell}
k^{\pm\ell}
\cr
&&
-{1\over4}\delta_{i[j_1}X_{j_2j_3]\ell_1\ell_2}\tilde{\omega}^{\pm\ell_1\ell_2}-{1\over36}\,{}^*S_{ij_1j_2j_3\ell_1\ell_2\ell_3}
\tilde{\varphi}^{\pm\ell_1\ell_2\ell_3}
\cr
&&
{\cal D}^{(\pm){\cal F}}_i\omega_{j_1j_2}^{\pm}\equiv \nabla_i \omega^\pm_{j_1j_2}\pm\partial_i \log A\omega^\pm_{j_1j_2}-{1\over2}X_{\ell_1\ell_2i[j_1}\varphi^{\pm\ell_1\ell_2}{}_{j_2]}
\cr
&&
=-{1\over4}X_{\ell_1\ell_2[ij_1}\varphi^{\pm\ell_1\ell_2}{}_{j_2]}
-{1\over18}\delta_{i[j_1}X_{j_2]\ell_1\ell_2\ell_3}\varphi^{\pm\ell_1\ell_2\ell_3}
\mp{1\over6}\,{}^*X_{ij_1j_2}\tilde{f}^\pm\pm{1\over3}S\delta_{i[j_1}\tilde{k}^\pm_{j_2]}~,~~~
\cr
&&
{\cal D}^{(\pm){\cal F}}_i \varphi_{j_1j_2j_3}^{\pm}\equiv \nabla_i\varphi^\pm_{j_1j_2j_3}\pm\partial_i\log A \varphi^\pm_{j_1j_2j_3}-{3\over2}X_{\ell i[j_1j_2}\omega^{\pm\ell}{}_{j_3]}\pm{3\over2}\,{}^*X_{i[j_1j_2}\tilde{k}^\pm_{j_3]}
\cr
&&
=\pm{4\over3}\,{}^*X_{[ij_1j_2}\tilde{k}^\pm_{j_3]}-{2\over3}X_{\ell[ij_1j_2}\omega^{\pm\ell}{}_{j_3]}
\pm\delta_{i[j_1}\,{}^*X_{j_2j_3]\ell}\tilde{k}^{\pm\ell}
\cr
&&
-{1\over4}\delta_{i[j_1}X_{j_2j_3]\ell_1\ell_2}\omega^{\pm\ell_1\ell_2}-{1\over36}\,{}^*S_{ij_1j_2j_3\ell_1\ell_2\ell_3}
\varphi^{\pm\ell_1\ell_2\ell_3}~,~~~
\cr
&&
{\cal D}^{(\pm){\cal F}}_i \tilde{f}^{\pm}\equiv \nabla_i\tilde{f}^\pm\pm\partial_i \log A\tilde{f}^\pm=\mp{1\over12}\,{}^*X_{i\ell_1\ell_2}\omega^{\pm\ell_1\ell_2}~,~~~
\cr
&&
{\cal D}^{(\pm){\cal F}}_i\tilde{k}_j^{\pm}\equiv \nabla_i\tilde{k}^\pm_j\pm\partial_i \log A\tilde{k}^\pm_j\pm{1\over4}\,{}^*X_{i\ell_1\ell_2}\varphi^{\pm\ell_1\ell_2}{}_{j}
\cr
&&
=\pm{1\over3}\,{}^*X_{\ell_1\ell_2[i}\varphi^{\pm\ell_1\ell_2}{}_{j]}\pm{1\over18}\delta_{ij}\,{}^*X_{\ell_1\ell_2\ell_3}
\varphi^{\pm\ell_1\ell_2\ell_3}\mp{1\over6}S\omega^\pm_{ij}~,~~~
\eea
where ${}^*X$ and ${}^*S$ are the Hodge duals of X and S on the internal space $M^7$, respectively,  and we have suppress the $r, s$ indices on the form bilinears that label the number of Killing spinors.

The action of the minimal TCFH connection on the space of forms preserves the subspaces of symmetric and skew-symmetric bilinears in the exchange of $\phi_\pm^r$ and $\phi_\pm^s$ Killing spinors and acts trivially on the scalars $f$ and $\tilde{f}$. As a consequence, the holonomy of the connection is included in $GL(63)\times GL(63)$. Note that the TCFH on the  0-, 1- and 3-form bilinears which are symmetric in the exchange of $\phi_\pm^r$ and $\phi_\pm^s$  is almost identical to that of the corresponding form bilinears which are skew-symmetric in the exchange of the same spinors.  The difference is a sign in the terms containing the fluxes  $S$ and ${}^*F$. The holonomy of the maximal TCFH connection is included in $GL(64)\times GL(64)$ since it acts non-trivially on the scalars.

The spinors $\sigma_\pm$ and $\tau_\pm$ are associated with the (reducible) Majorana representation $\Delta_{\bf 16}$ of $\mathfrak{spin}(8)$. This decomposes under $\mathfrak{spin}(7)$ as $\Delta_{\bf 16}=\Delta_{\bf 8}\oplus \Delta_{\bf 8}$, where $\Delta_{\bf 8}$ is the (irreducible) Majorana representation  of $\mathfrak{spin}(7)$.   Moreover the tensor product of two such representations, $\Delta_{\bf{8}}$, decomposes in terms of form representations as

\bea
\Delta_{\bf{8}} \otimes \Delta_{\bf{8}}=\sum_{k=0}^{3} \Lambda^k(\mathbb{R}^7)~.
\eea
Clearly the TCFH includes of two copies of forms that appear in the above decomposition which lead for the holonomy of the maximal connection of TCFH to be included in $GL(64)\times GL(64)$.

\section{The TCFH of warped AdS$_k$, $k=5,6, 7$,  backgrounds}

\subsection{Fields and Killing spinors}

The fields of 11-dimensional supergravity for warped AdS$_k\times_w M^{11-k}$, $k=5,6,7$, backgrounds can be written as
\bea
g = 2 \bbe^+ \bbe^-+ (\bbe^z)^2+\sum_a (\bbe^a)^2+ g(M^{11-k})~,~~~F=X~,
 \label{metr}
\eea
with $g(M^{11-k})= \delta_{ij} \bbe^i \bbe^j$, where $X$ is a 4-form on the internal space $M^{11-k}$.  The null pseudo-orthonormal frame $(\bbe^+, \bbe^-, \bbe^z, \bbe^a, \bbe^i)$ is expressed as
\bea
\bbe^+=du~,~~~\bbe^-=dr - 2 \ell^{-1} r dz - 2 r d \ln A~,~~~\bbe^z= A dz~,~~~\bbe^a= A e^{{z\over\ell}} dx^a~,~~~\bbe^i=e^i_J dy^j~,
\eea
where $y$ are coordinates of the internal space $M^{11-k}$ and  $(u,r,z,x^a)$ are the rest of the coordinates of the spacetime, $\ell$ is the radius of AdS subspace   and $A$ is the warp factor.

As in all previous cases, the KSEs of 11-dimensional supergravity can be integrated  over the AdS subspace and the Killing spinors can be expressed as, $\epsilon=\epsilon(\sigma_\pm)+\epsilon(\tau_\pm)$, see \cite{Mads},
with
\bea
\epsilon(\sigma_\pm)&=&\sigma_++\sigma_--\ell^{-1} e^{{z\over\ell}} x^a \Gamma_{az} \sigma_--\ell^{-1} A^{-1} u \Gamma_{+z} \sigma_-~,
\cr
\epsilon(\tau_\pm)&=&e^{-{z\over\ell}}\tau_+-\ell^{-1} A^{-1} r e^{-{z\over\ell}} \Gamma_{-z} \tau_+-\ell^{-1} x^a \Gamma_{az} \tau_++e^{{z\over \ell}} \tau_-~,
\label{kspadsk2}
\eea
where the $\sigma_\pm$ and $\tau_\pm$  spinors satisfy the lightcone projections $\Gamma_\pm\sigma_\pm=\Gamma_\pm\tau_\pm=0$, and depend only on the coordinates of $M^{11-k}$.  In addition, they satisfy the KSEs
\bea
{\cal D}^{\pm}_i \phi_\pm=0~,
\eea
along the internal space $M^{11-k}$, where the supercovariant connection is
\bea
{\cal D}^{\pm}_i \equiv \nabla_i \pm {1\over 2}\partial_i \log A-{1\over288}\sgX_i+{1\over36}\sX_i~,
\eea
$\nabla$ is the connection on the spin bundle of $M^{11-k}$ induced from the metric $g(M^{11-k})$ and $\phi_\pm$ stands from either $\sigma_\pm$ or $\tau_\pm$. Note that for warped AdS$_7$ backgrounds the term $\sgX$ in the supercovariant connection vanishes. As in previous cases $\sigma_\pm$ and $\tau_\pm$ satisfy an additional algebraic KSE  which arises from the integration of the gravitino KSE of 11-dimensional supergravity along the $z$ coordinate and can be found in \cite{Mads}.  It will be used to determine  the number of Killing spinors
in some examples below.

\subsection{The TCFH of warped AdS$_5$ backgrounds}

A basis in the space of form bilinears on the internal space $M^6$ is
\bea\label{ads5 symm}
&&f^{\pm r,s}=\langle \phi_\pm^r, \phi_\pm^s \rangle~,~~~
k^{\pm r,s}=\langle \phi_\pm^r, \Gamma_i\phi_\pm^s\rangle\, \bbe^i~,~~~
\tilde{k}^{\pm r,s}=\langle \phi_\pm^r, \Gamma_i\Gamma_{(3)}\phi_\pm^s\rangle\, \bbe^i ~,~~~
\cr
&&
\tilde{\omega}^{\pm r,s}={1\over2}\langle \phi_\pm^r, \Gamma_{ij}\Gamma_{(3)}\phi_\pm^s\rangle\, \bbe^i \wedge \bbe^j~,~~~
\omega^{\pm r,s}={1\over2}\langle \phi_\pm^r, \Gamma_{ij}\phi_\pm^s\rangle\, \bbe^i\wedge \bbe^j~,~~~
\cr
&&
\varphi_{ijk}^{\pm r,s}={1\over 3!}\langle \phi_\pm^r, \Gamma_{ijk}\phi_\pm^s\rangle\, \bbe^i\wedge \bbe^j \wedge \bbe^k~,~~~
\tilde{f}^{\pm r,s}=\langle \phi_\pm^r,\Gamma_{(3)} \phi_\pm^s \rangle~,
\eea
where $\Gamma_{(3)}=\Gamma_{zx^1x^2}$, i.e. it is the product of gamma matrices along the directions $\bbe^z$ and $\bbe^a$ for $a=1,2$.  The first four bilinears
are symmetric in the exchange of $\phi_\pm^r$ and $\phi_\pm^s$ spinors while the rest are skew-symmetric.

The TCFH expressed\footnote{In our conventions $\Gamma_{(3)}\Gamma_{i_1\dots i_6}\phi_{\pm}=\pm\epsilon_{i_1\dots i_6}\phi_{\pm}$ and $\epsilon^{123456}=1$.} in terms of the minimal connection is
\bea
\label{tcfh ads5-1}
&&
{\cal D}^{(\pm)\cal F}_i f^{\pm} \equiv \nabla_i f^\pm\pm\partial_i \log A f^\pm=\pm{1\over6} \,{}^*X_{i\ell}\tilde{k}^\pm{}^{\ell}~,~~~
\cr
&&
{\cal D}^{(\pm)\cal F}_i k_j^{\pm}\equiv \nabla_i k^\pm_j\pm\partial_i\log A k^\pm_j\pm{1\over2}\,{}^*X_{i\ell}\tilde{\omega}^\pm{}^{\ell}{}_j
\cr
&&
=\pm{2\over3}{}^*X_{[i|\ell|}\tilde{\omega}^\pm{}^{\ell}{}_{j]}\mp{1\over6}\delta_{ij}\,{}^*X_{\ell_1\ell_2}\tilde{\omega}^\pm{}^{\ell_1\ell_2}~,~~~
\cr
&&
{\cal D}^{(\pm)\cal F}_i\tilde{k}_j^{\pm}\equiv \nabla_i \tilde{k}^\pm_j\pm\partial_i \log A\tilde{k}^\pm_j={1\over6}X_{ij\ell_1\ell_2}\tilde{\omega}^\pm{}^{\ell_1\ell_2}\pm{1\over6}\,{}^*X_{ij}f^\pm~,~~~
\cr
&&
{\cal D}^{(\pm)\cal F}_i \tilde{\omega}_{j_1j_2}^{\pm}\equiv \nabla_i \tilde{\omega}^\pm_{j_1j_2}\pm\partial_i \log A\tilde{\omega}^\pm_{j_1j_2}\pm\,{}^*X_{i[j_1}k^\pm_{j_2]}
\cr
&&
=-{1\over3}X_{ij_1j_2\ell}\tilde{k}^\pm{}^{\ell}
\pm{2\over3}\delta_{i[j_1}\,{}^*X_{j_2]\ell}k^\pm{}^\ell\pm\,{}^*X_{[ij_1}k^\pm_{j_2]}~,
\cr
&&
{\cal D}^{(\pm)\cal F}_i\omega_{j_1j_2}^{\pm}\equiv \nabla_i \omega^\pm_{j_1j_2}\pm\partial_i\log A \omega^\pm_{j_1j_2}-{1\over2}X_{\ell_1\ell_2i[j_1}\varphi^\pm{}^{\ell_1\ell_2}{}_{j_2]}
\cr
&&
=-{1\over4}X_{\ell_1\ell_2[ij_1}\varphi^\pm{}^{\ell_1\ell_2}{}_{j_2]}-{1\over18}\delta_{i[j_1}X_{j_2]\ell_1\ell_2\ell_3}\varphi^\pm{}^{\ell_1\ell_2\ell_3}~,~~~
\cr
&&
{\cal D}^{(\pm)\cal F}_i \varphi_{\ell_1\ell_2\ell_3}^{\pm}\equiv \nabla_i \varphi^\pm_{\ell_1\ell_2\ell_3}\pm\partial_i \log A \varphi^\pm_{\ell_1\ell_2\ell_3}-{3\over2}X_{\ell i[j_1j_2}\omega^\pm{}^{\ell}{}_{j_3]}
\cr
&&
=-{1\over4}\delta_{i[j_1}X_{j_2j_3]\ell_1\ell_2}\omega^\pm{}^{\ell_1\ell_2}-{2\over3}X_{\ell[ij_1j_2}\omega^\pm{}^{\ell}{}_{j_3]}
\pm\delta_{i[j_1}\,{}^*X_{j_2j_3]}\tilde{f}^\pm~,~~~
\cr
&&
{\cal D}^{(\pm)\cal F}_i\tilde{f}^{\pm}\equiv \nabla_i {\tilde{f}}^\pm\pm\partial_i\log A\tilde{f}^\pm=\mp{1\over6}\,{}^*X_{\ell_1\ell_2}\varphi^\pm{}^{\ell_1\ell_2}{}_{i}~,
\eea
where ${}^*X$ is the Hodge dual of $X$ on $M^6$ and we have suppressed the indices $r,s$ of the form bilinears.

As expected the minimal connection of the TCFH ${\cal D}^{(\pm)\cal F}$  is not form degree preserving. On the other hand its action closes on the form bilinears which are either symmetric or skew-symmetric in the interchange of spinors $\phi_\pm^r$ and $\phi_\pm^s$.  The holonomy of ${\cal D}^{(\pm)\cal F}$  is contained in $SO(6)\times GL(21)\times GL(35)$ as in addition it acts with the Levi-Civita connection on the 1-form bilinear $A\, \tilde k$ and trivially on the scalar bilinears $f$ and $\tilde f$.  Note that the holonomy of the maximal TCFH connection is contained in $GL(28)\times GL(35)$ as it only closes on the symmetric and skew-symmetric form bilinears under the exchange of spinors $\phi_\pm^r$ and $\phi_\pm^s$.

\subsection{ The TCFH of warped AdS$_6$  backgrounds}

A basis in the space of form bilinears on the internal space $M^5$ is
\bea
\label{ads6 symm}
&&f^{\pm r,s}=\langle \phi_\pm^r, \phi_\pm^s \rangle~,~~~
k^{\pm r,s}=\langle \phi_\pm^r, \Gamma_i\phi_\pm^s\rangle\, \bbe^i~,~~~
\tilde{f}^{\pm r,s}=\langle \phi_\pm^r,\Gamma_{(4)} \phi_\pm^s \rangle~,~~~
\cr
&&
\tilde{k}^{\pm r,s}=\langle \phi_\pm^r, \Gamma_i\Gamma_{(4)}\phi_\pm^s\rangle\, \bbe^i~,~~~
\cr
&&\omega^{\pm r,s}={1\over2}\langle \phi_\pm^r, \Gamma_{ij}\phi_\pm^s\rangle\, \bbe^i \wedge \bbe^j\,~,~~~
\tilde{\omega}^{\pm r,s}={1\over2}\langle \phi_\pm^r, \Gamma_{ij}\Gamma_{(4)}\phi_\pm^s\rangle\, \bbe^i\wedge \bbe^j ~,~~~
\eea
where the first four form bilinears are symmetric in the exchange of $\phi_\pm^r$ and $\phi_\pm^s$ spinors while the rest are skew-symmetric and  $\Gamma_{(4)}=\Gamma_{zx^1x^2x^3}$.

The TCFH  expressed\footnote{In our conventions $\Gamma_{(4)}\Gamma_{i_1\dots i_5}\phi_{\pm}=\pm\epsilon_{i_1\dots i_5}\phi_{\pm}$ with $\epsilon^{12345}=1$.}  in terms of the minimal connection, ${\cal D}^{(\pm)\cal F}$, is
\bea
&&
{\cal D}^{(\pm)\cal F}_i f^{\pm}\equiv \nabla_i f^\pm\pm\partial_i \log A f^\pm=\pm{1\over6}\,{}^*X_i\tilde{f}^\pm~,
\cr
&&
{\cal D}^{(\pm)\cal F}_i k_j^{\pm}\equiv \nabla_i k^\pm_j\pm\partial_i\log A k^\pm_j\mp{1\over2}\,{}^*X_i\tilde{k}^\pm_j=\mp{2\over3}\,{}^*X_{[i}\tilde{k}^\pm_{j]}\mp{1\over3}\delta_{ij}\,{}^*X_{\ell}\tilde{k}^\pm{}^{\ell}~,~~~
\cr
&&
{\cal D}^{(\pm)\cal F}_i \tilde{f}^{\pm} \equiv \nabla_i \tilde{f}^\pm\pm\partial_i \log A \tilde{f}^\pm=\pm{1\over6}\,{}^*X_i f^\pm~,~~~
\cr
&&
{\cal D}^{(\pm)\cal F}_i \tilde{k}_j^{\pm}\equiv \nabla_i \tilde{k}^\pm_j\pm\partial_i \log A \tilde{k}^\pm_j\mp{1\over2}\,{}^*X_i k^\pm_j=\mp{2\over3}\,{}^*X_{[i}k^\pm_{j]}\mp{1\over3}\delta_{ij}\,{}^*X_\ell k^\pm{}^\ell~,
\cr
&&
{\cal D}^{(\pm)\cal F}_i\omega_{j_1j_2}^{\pm}\equiv \nabla_i \omega^\pm_{j_1j_2}\pm\partial_i\log A \omega^\pm_{j_1j_2}\mp{1\over2}\,{}^*X_i\tilde{\omega}^\pm_{j_1j_2}=\mp\,{}^*X_{[i}\tilde{\omega}^\pm_{j_1j_2]}
\mp{2\over3}\delta_{i[j_1}\tilde{\omega}^\pm{}^{\ell}{}_{j_2]}\,{}^*X_\ell~,~~~
\cr
&&
{\cal D}^{(\pm)\cal F}_i \tilde{\omega}_{j_1j_2}^{\pm}\equiv \nabla_i \tilde{\omega}^\pm_{j_1j_2}\pm\partial_i \log A \tilde{\omega}^\pm_{j_1j_2}\mp{1\over2}\,{}^*X_i\omega^\pm_{j_1j_2}=\mp\,{}^*X_{[i}\omega^\pm_{j_1j_2]}\mp{2\over3} \delta_{i[j_1}\omega^\pm{}^\ell{}_{j_2]}\,{}^*X_{\ell}~,
\eea
where ${}^*X$ is the Hodge dual of $X$ and we have suppressed the $r,s$ indices on the form bilinears as in previous cases.

Unlike previous cases, the minimal TCFH connection  ${\cal D}^{(\pm)\cal F}$ for AdS$_6$ backgrounds is form  degree preserving. Furthermore its action can be diagonalised on the forms
\bea
\zeta^\pm_{(+)}=\zeta^\pm+ \tilde \zeta^\pm~,~~~\zeta^\pm_{(-)}=\zeta^\pm- \tilde \zeta^\pm~,
\eea
where $\zeta^\pm$ stands for either $k^\pm$ or $\omega^\pm$, i.e. one has that
\bea
&&{\cal D}^{(\pm)\cal F}_i \zeta^\pm_{(+)}=\nabla_i \zeta^\pm_{(+)}\pm\partial_i \log A \zeta^\pm_{(+)} \mp{1\over2}\,{}^*X_i \zeta^\pm_{(+)}~,
\cr
&&{\cal D}^{(\pm)\cal F}_i \zeta^\pm_{(-)}=\nabla_i \zeta^\pm_{(-)}\pm\partial_i \log A \zeta^\pm_{(-)} \pm{1\over2}\,{}^*X_i \zeta^\pm_{(-)}~.
\eea
Such a connection arises provided one gauges the scale transformation $\zeta\rightarrow s\,  \zeta $ accompanied with
${}^*X\rightarrow {}^* X\pm 2 s^{- 1} d s$, where the sign is plus for $\zeta^+_{(+)}$ and $\zeta^-_{(-)}$ while it is minus for the rest of the form bilinears. Clearly, there are two sectors and the holonomy of the connection in each sector is $SO(5)\times (\bR-\{0\})$.

\subsection{The TCFH of AdS$_7$ backgrounds}

A basis in the space of spinor bilinears on the internal space $M^4$ is
\bea
\label{ads7 symm}
&&
f^{\pm r,s}=\langle \phi_\pm^r, \phi_\pm^s \rangle~,~~~
k^{\pm r,s}=\langle \phi_\pm^r, \Gamma_i\phi_\pm^s\rangle\, \bbe^i~,~~~
\tilde{f}^{\pm r,s}=\langle \phi_\pm^r,\Gamma_{(5)} \phi_\pm^s \rangle~,~~~
\cr
&&
\tilde{k}^{\pm r,s}=\langle \phi_\pm^r, \Gamma_i\Gamma_{(5)}\phi_\pm^s\rangle\, \bbe^i ~,~~~
\omega^{\pm r,s}={1\over2} \langle \phi_\pm^r, \Gamma_{ij}\phi_\pm^s\rangle\, \bbe^i \wedge \bbe^j~,
\eea
where the first three are symmetric in the exchange of spinors $\phi^r_\pm$ and $\phi^s_\pm$ while the rest are skew-symmetric and $\Gamma_{(5)}=\Gamma_{zx^1x^2x^3x^4}$.

The TCFH expressed\footnote{In our conventions $\Gamma_{(5)}\Gamma_{i_1\dots i_4}\phi_{\pm}=\pm\epsilon_{i_1\dots i_4}\phi_{\pm}$ and $\epsilon^{1234}=1$.}   in terms of the minimal connection ${\cal D}^{(\pm)\cal F}$ is

\bea
&&
{\cal D}^{(\pm)\cal F}_i f^{\pm}\equiv \nabla_i f^\pm \pm\partial_i \log A f^\pm =0~,~~~
\cr
&&
{\cal D}^{(\pm)\cal F}_i k_j^{\pm}\equiv \nabla_i k^\pm_j\pm\partial_i\log A k^\pm_j=\mp{1\over3}\delta_{ij}\,{}^* X\tilde{f}^\pm~,~~~
\cr
&&
{\cal D}^{(\pm)\cal F}_i \tilde{f}^{\pm}\equiv \nabla_i \tilde{f}^\pm \pm\partial_i \log A \bar{f}^\pm =\pm{1\over3}\,{}^*Xk^\pm _i~,
\cr
&&
{\cal D}^{(\pm)\cal F}_i\omega_{j_1j_2}^{\pm}\equiv \nabla_i \omega^\pm_{j_1j_2}\pm\partial_i\log A\omega^\pm_{j_1j_2}=\pm{2\over3}\delta_{i[j_1}\bar{k}^\pm_{j_2]}\,{}^*X~,~~~
\cr
&&
{\cal D}^{(\pm)\cal F}_i \tilde{k}_j^{\pm}\equiv \nabla_i \tilde{k}^\pm_j\pm\partial_i\log A\tilde{k}^\pm_j=\mp{1\over3}\,{}^*X\omega^\pm_{ij}~,
\label{tcfhads7}
\eea
where ${}^*X$ is the Hodge dual of the 4-form $X$ on the internal space $M^4$.

It is clear that the (reduced) holonomy of   ${\cal D}^{(\pm)\cal F}$ is contained in $SO(4)$. Furthermore  $A^{\pm 1} k^\pm$ and $A^{\pm 1} \omega^\pm$ are CCKY forms.  Therefore their dual in $M^4$ are KY.  In addition $A^{\pm 1} \tilde k^\pm$ is a KY tenors. It is well-known KY tensors generate symmetries in spinning particle actions.

\section{Probes and symmetries} \label{prob}

\subsection{Relativistic and spinning particles}

We have integrated the KSE of 11-dimensional supergravity along the AdS$_k$ subspace of a warped spacetime, AdS$_k\times_w M^{11-k}$, and found the TCFHs on the
internal space $M^{11-k}$.  To investigate whether the form bilinears of the TCFHs on the internal space generate symmetries for spinning particle actions, we have to integrate the dynamics of the spinning particle along the AdS$_k$ subspace and describe the effective dynamics of the system on the internal space $M^{11-k}$.

For this consider first the dynamics of a relativistic particle on a warped spacetime, $N\times_w M$, with metric $g=A^2 h+ \gamma$, where $h$
is a metric on $N$ and $\gamma$ is a metric on $M$ and $A$ is the warped factor. Varying the action
\bea
A={1\over2} \int\, dt\, g_{\mu\nu} \dot x^\mu \dot x^\nu~,
\eea
one finds that the equations of motion are
\bea
\nabla^h_t (A^2 \dot \rho^a)=0~,~~~\nabla_t^\gamma \dot y^I-{1\over2} \gamma^{IJ} \partial_J A^2\, h_{ab} \dot\rho^a \dot\rho^b=0~,
\eea
where $\nabla^h$ and $\nabla^\gamma$ denote the Levi-Civita connections of $h$ and $\gamma$, respectively, $\rho^a$ are coordinates on $N$ and $y^I$ are coordinates on $M$.  It is clear that
\bea
Q^2={1\over2} A^4 h_{ab} \dot\rho^a \dot\rho^b~,
\eea
is  conserved as a consequence of the field equation on $N$.  Then notice that the dynamics of the relativistic particle  on $M$ can be described by the
effective action
\bea
A_M={1\over2}\int dt\, \big( {1\over2} \gamma_{IJ} \dot y^I \dot y^J-Q^2 A^{-2}\big)~.
\label{fdm}
\eea
The action apart from the usual kinetic term exhibits a potential depending on the warped factor. There are various sectors to consider parameterised by the value of $Q^2$. If either $Q^2=0$, which is  the case for $\rho$ constant, or $A^2$ is constant, $A_M$ becomes the standard action for geodesic motion on $M$ possibly shifted by an ignorable constant.

A similar analysis can be performed for a spinning particle probe \cite{bvh} propagating on a spacetime  with metric $g$  described by the action
\bea
A=-{i\over2} \int\, dt\, d\theta\,\, g_{\mu\nu}\, D x^\mu\, \dot x^\nu~,
\label{sgact}
\eea
where $(t,\theta)$ are  worldline superspace coordinates,    $x$ are worldline superfields $x=x(t, \theta)$ and $D$ is a worldline superspace derivative with  $D^2=i\partial_t$.

The equations of motion of the spinning particle (\ref{sgact}) propagating on a warped spacetime $N\times_w M$, as for the relativistic particle above, are
\bea
\nabla^h (A^2 \dot \rho^a)+\nabla^h_t (A^2 D\rho^a)=0~,~~~\nabla^\gamma \dot y^I={1\over2} \gamma^{IJ} \partial_J A^2 h_{ab} D\rho^a \dot \rho^b~.
\eea
In this case, there is not a simple description of the effective dynamics on $M$ as for the relativistic particle described by the action (\ref{fdm}). However note that $\rho$ equals to a constant is a solution of the equations of motion above.  So if either $\rho$ is constant or the warp factor $A$ is constant, the effective
dynamics of the spinning particle on $M$ is described by the action
\bea
A_M=-{i\over2} \int\, dt\, d\theta\,\, \gamma_{IJ}\, D y^I\, \dot y^J~.
\label{espm}
\eea
It is well known that the action above is invariant under an infinitesimal transformation
\bea
\delta y^I=\epsilon\, \alpha^I{}_{J_1\cdots J_{m-1}} Dy^{J_1}\cdots Dy^{J_{m-1}}~,
\label{svar}
\eea
provided that $\alpha$ is a KY form \cite{gibbons}, where $\epsilon$  is the infinitesimal parameter.

There is an extensive list of 11-dimensional supersymmetric AdS solutions, see e.g. \cite{mads2a, mads23, mads4, mads5, mads6a, mads6b}. The purpose here is to give some examples of TCFHs and investigate their properties instead of being comprehensive. So we shall focus below on the TCFH of the maximally supersymmetric AdS solutions and some AdS solutions that arise as near horizon geometries of intersecting M-branes.

\subsection{Maximally supersymmetric AdS backgrounds}

\subsubsection{AdS$_4 \times S^7$}

The TCFH of warped AdS$_4$ backgrounds with only electric flux, i.e. $X=0$ in (\ref{fads4}), can be written as
\bea
&&
{\cal D}^{(\pm)\cal F}_i f^{\pm}\equiv \nabla_i f^\pm\pm\partial_i \log A\, f^\pm=0~,~~~
\cr
&&
{\cal D}^{(\pm)\cal F}_i k_j^{\pm}\equiv \nabla_i k^\pm _j\pm\partial_i\log A\, k^\pm_j
=\pm{1\over6}S\tilde{\omega}^\pm_{ij}~,~~~
\cr
&&
{\cal D}^{(\pm)\cal F}_i\tilde{\omega}_{j_1j_2}^{\pm}\equiv \nabla_i\tilde{\omega}^\pm_{j_1j_2}\pm\partial_i \log A\, \tilde{\omega}^\pm_{j_1j_2}
=\mp{1\over3}S\delta_{i[j_1}k^\pm_{j_2]}~,~~~
\cr
&&
{\cal D}^{(\pm)\cal F}_i \tilde{\varphi}_{j_1j_2j_3}^{\pm}\equiv \nabla_i \tilde{\varphi}^\pm_{j_1j_2j_3}\pm\partial_i \log A\, \tilde{\varphi}^\pm_{j_1j_2j_3}
=-{1\over36}\,{}^*S_{ij_1j_2j_3\ell_1\ell_2\ell_3}\tilde{\varphi}^\pm{}^{\ell_1\ell_2\ell_3}
\cr
&&
{\cal D}^{(\pm)\cal F}_i\omega_{j_1j_2}^{\pm}\equiv \nabla_i \omega^\pm_{j_1j_2}\pm\partial_i \log A\,\omega^\pm_{j_1j_2}
=\pm{1\over3}S\delta_{i[j_1}\tilde{k}^\pm_{j_2]}~,~~~
\cr
&&
{\cal D}^{(\pm)\cal F}_i \varphi_{j_1j_2j_3}^{\pm}\equiv \nabla_i\varphi^\pm_{j_1j_2j_3}\pm\partial_i\log A\, \varphi^\pm_{j_1j_2j_3}
=-{1\over36}\,{}^*S_{ij_1j_2j_3\ell_1\ell_2\ell_3}\varphi^\pm{}^{\ell_1\ell_2\ell_3}~,~~~
\cr
&&
{\cal D}^{(\pm)\cal F}_i \tilde{f}^{\pm}\equiv \nabla_i\tilde{f}\pm\partial_i \log A\, \tilde{f}^\pm=0~,~~~
\cr
&&
{\cal D}^{(\pm)\cal F}_i\tilde{k}_j^{\pm}\equiv \nabla_i\tilde{k}^\pm_j\pm\partial_i \log A\, \tilde{k}^\pm_j
=\mp{1\over6}S\omega^\pm_{ij}~.~~~
\eea
It is clear from the TCFH above that $A^{\pm1} k$,  $A^{\pm1} \tilde k$, $A^{\pm1} \varphi$ and $A^{\pm1} \tilde\varphi$ are KY forms, and $A^{\pm1} \omega$
and $A^{\pm1} \tilde\omega$ are CCKY forms. As $A^{\pm1} \omega$
and $A^{\pm1} \tilde\omega$ are CCKY forms their duals  in the internal space are KY forms.  The holonomy of the TCFH connection is included in $SO(7)$. 

 The  maximally supersymmetric AdS$_4$ solution is a Freund-Rubin type of background with internal space $M^7$  the round 7-sphere, $M^7=S^7$, and the warp factor $A$ constant.  All the forms bilinears above generate symmetries for the spinning particle probe action (\ref{espm}). Note that the form bilinears on $S^7$ are not necessarily invariant forms under the $SO(8)$ isometry group of $S^7$.

\subsubsection{AdS$_7 \times S^4$}

The maximally supersymmetric AdS$_7$ solution is again a Freund-Rubin type of background with internal space $M^4$  the round 4-sphere, $M^4=S^4$, and the warp factor $A$  constant.  An inspection of the TCFH of warped AdS$_7$ backgrounds in (\ref{tcfhads7}) reveals that the bilinear
$\tilde k^\pm$ is a KY form, and $k^\pm$ and $\omega^\pm$ are CCKY forms. Again the duals ${}^*k^\pm$ and  ${}^*\omega^\pm$ of  $k^\pm$ and $\omega^\pm$ in $S^4$, respectively, are KY forms and so $\tilde k^\pm$, ${}^*k^\pm$ and  ${}^*\omega^\pm$ generate symmetries for the spinning particle probe action  (\ref{espm}).

\subsection{AdS backgrounds from intersecting branes}

More examples of AdS backgrounds emerge as near horizon geometries of intersecting M-branes \cite{ptgp}. We shall not explore all the possibilities, see  \cite{bps} for more examples. Instead we shall focus on the
AdS$_2$ and  AdS$_3$ solutions that arise as near horizon geometries of the intersection of three M2-branes on an 0-brane, the intersection of an M2-brane and M5-brane on a 1-brane and the intersection of three M5-branes on a 1-brane configurations.

 \subsubsection{AdS$_2$ solution from intersecting M2-branes}

 One could take the near horizon geometry of the three intersecting M2-brane solution on a 0-brane and proceed to examine the associated TCFH. Instead, we shall write an ansatz
 for the fields which includes the solution.  In particular, we set
 \bea
 g=g_\ell(AdS_2)+ g(S^3)+ g(\bR^6)~,~~~F=d\mathrm{vol}_\ell(AdS_2)\wedge Y~,
 \label{antz1}
 \eea
  where $g_\ell(AdS_2)$ and $d\mathrm{vol}_\ell(AdS_2)$ is the standard metric and volume 2-form on AdS$_2$ with radius $\ell$, respectively, $g(S^3)$ is the round metric on $S^3$ of unit radius, $g(\bR^6)$ is the Euclidean metric on $\bR^6$ and $Y$ is a constant 2-form on $\bR^6$.  Using the scale symmetry $g\rightarrow \Omega^2 g$ and $F\rightarrow \Omega^3 F$ of 11-dimensional supergravity as well as some coordinate transformations, one can show that the near horizon geometry  of three M2-branes, with arbitrary charge densities, intersecting on a 0-brane solution can be cast into the above form. Clearly for this ansatz  $X=0$ in (\ref{ads2f}) and we have  set $A=1$.

 Focusing on the $\phi_+$ Killing spinors, the gravitino KSE along the directions of $\bR^6$ implies that $Y$ is a non-degenerate 2-form and proportional
 to a K\"ahler form $\lambda$ on $\bR^6$ associated with the Euclidean metric, i.e. $Y=\gamma \lambda$, $\gamma\in \bR$.  Furthermore $\phi_+$ has to satisfy the conditions
  $\Gamma_{1234}\phi_+=\eta \phi_+$ and $\Gamma_{1256}\phi_+=\zeta \phi_+$, where $\eta, \zeta=\pm 1$ and we have arranged such that $\bR^6$ lies in the directions $1, \dots, 6$. Then the warp factor field equation in \cite{Mads} implies that
 $\gamma^2=\ell^{-2}$. Next the field equation along $S^3$, which is the round unit sphere, gives $\ell^{-2}=4$. All the remaining KSEs and field equations are satisfied without further conditions. Therefore the background (\ref{antz1}) with
 the above choice of parameters  admits 4 $\phi_+$ Killing spinors.  The solution also admits  4 more  $\phi_-$ Killing spinor  and so it preserves $1/4$ of the supersymmetry.

 Next notice that the supercovariant derivative along $S^3$ can be written as
 \bea
 {\cal D}_\alpha^{\pm}=\nabla^{S^3}_\alpha\pm {1\over 24} \Gamma_\alpha \slashed {Y}~,
 \eea
 where $\alpha$ here labels the three orthonormal directions tangential to $S^3$.  Note that $\Gamma_\alpha \slashed {Y}= \slashed {Y} \Gamma_\alpha$.  Moreover  considering only those components of the form bilinears that lie on $S^3$, i.e.
 \bea
 k^{\pm rs}_\alpha=\langle\phi^r_\pm, \Gamma_\alpha\phi^s_\pm\rangle~,~~~\omega^{\pm rs}_{\alpha\beta}=\langle\phi^r_\pm, \Gamma_{\alpha\beta}\phi^s_\pm\rangle~,~~~\varphi^{\pm rs}_{\alpha\beta\delta}=
 \langle\phi^r_\pm, \Gamma_{\alpha\beta\delta}\phi^s_\pm\rangle~,
 \label{s3bi}
 \eea
 one can demonstrate that $k$ and $\varphi$ are KY forms while $\omega$ is CCKY form.  Therefore all of them or their Hodge duals on $S^3$ generate symmetries for the probe action
 (\ref{espm}).

 \subsubsection{AdS$_3$ solution from M2- and  M5-branes}

An ansatz which includes the near horizon geometry of an M2-brane intersecting an M5-brane on a 1-brane is
\bea
g=g_\ell (AdS_3)+ g(S^3)+ g(\bR^5)~,~~~F= d\mathrm{vol}_\ell(AdS_3) \wedge Q+   d\mathrm{vol}(S^3)\wedge P~,
\eea
i.e. $Q,X\not=0$ in (\ref{ads3f}), where $g_\ell (AdS_3)$ and $d\mathrm{vol}_\ell(AdS_3)$ are the standard metric and volume 3-form of AdS$_3$ with radius $\ell$, respectively, and $g(\bR^5)$ is the Euclidean metric
on $\bR^5$.  Similarly,  $g(S^3)$ and $d\mathrm{vol}(S^3)$ are the metric and volume 3-form of unit round 3-sphere, respectively,  and the 1-forms $P$ and $Q$ are constant and lie along the same direction in $\bR^5$, e.g. $P=p\, dw$ and $Q=q\, dw$ with $p,q$ constants.

Focusing on the KSEs on $\sigma_+$ and setting without loss of generality $A=1$, the integrability of the gravitino KSE along the $\bR^5$ directions implies that $p^2=q^2$. Moreover, one has to also consider the algebraic KSE $\Xi^+ \sigma_+=0$ which arises from the integration of the KSE of 11-dimensional supergravity along the $z$ direction of AdS$_3$. As $\Xi^+=-(2\ell)^{-1}+{1\over 288} \Gamma_z \slashed{X}+ {1\over 6} \slashed {Q}$, see \cite{Mads}, the algebraic KSE can be arranged as
\bea
\Big( -{1\over \ell} \Gamma_w + {p\over 6} \Gamma_z \Gamma_{(3)}+{1\over3} q\Big)\sigma_+=0~,
\label{ads3proj}
\eea
where $\Gamma_{(3)}$ is the product of the three gamma matrices along orthonormal directions tangential to $S^3$.

As $p^2=q^2$ to solve (\ref{ads3proj})    let us set $p=q$. The other case $p=-q$ can be treated in a similar way. Then decompose (\ref{ads3proj}) into eigenspaces
of $\Gamma_w$ and $\Gamma_z\Gamma_{(3)}$.  Using $\Gamma_w^2=1$ and $(\Gamma_z\Gamma_{(3)})^2=1$ and $\Gamma_w \Gamma_z\Gamma_{(3)}=\Gamma_z\Gamma_{(3)} \Gamma_w$, we have that
\bea
- \eta {1\over \ell}  + \zeta {q\over 6} +{1\over3} q=0~,
\eea
where
$\Gamma_w\sigma_+=\eta \sigma_+$ and $\Gamma_z\Gamma_{(3)}\sigma_+=\zeta \sigma_+$ with $\eta, \zeta=\pm1$. There are four cases to consider leading to
  $q=\pm 2\ell^{-1}$ and $q=\pm 6 \ell^{-1}$.  Two of these solutions are related to the other two by a change of the overall sign of the 4-form field strength $F$. So there are only two remaining independent solutions.  Furthermore the $q=\pm 6 \ell^{-1}$ solution is ruled out by the warp factor field equation \cite{Mads}. In addition, the field equation along $S^3$ implies that $p^2=4$.  As  $q=\pm 2\ell^{-1}$, one finds that $\ell^2=1$, which is the near horizon
geometry of the M2- and M5-brane intersection on a 1-brane solution. This solution preserves 1/2 of supersymmetry as each of the KSEs on $\sigma_\pm$ and $\tau_\pm$ give 4 independent solutions.

As in the previous AdS$_2$ case, we next consider the KSE along the $S^3$ directions whose supercovariant derivative can be put in the form
\bea
{\cal D}_\alpha^+=\nabla^{S^3}_\alpha+\Gamma_\alpha \Gamma_z ({1\over6} p \zeta+{1\over 12} q) \eta~,
\eea
where $\alpha$ labels the three othonormal directions tangential to $S^3$.  Considering the form bilinears (\ref{s3bi}) with $\phi_+=\sigma_+$, it is easy to show that $k$ and $\omega$ are CCKY forms on $S^3$.  As a result their Hodge duals on $S^3$ are KY forms and generate symmetries for the probe action (\ref{espm}). The bilinear $\varphi$ is also a CCKY form but its dual is a scalar.

\subsection{AdS$_3$ solution for intersecting M5-branes}

An ansatz which includes the near horizon geometry of three M5-branes intersecting  on a 1-brane is
\bea
g=g_\ell (AdS_3)+ g(S^2)+ g(\bR^6)~,~~~F=d\mathrm{vol}(S^2)\wedge W~,
\eea
i.e. $Q=0$ in (\ref{ads3f}), where $g_\ell (AdS_3)$ is the metric of AdS$_3$ with radius $\ell$, $g(S^2)$ and $d\mathrm{vol}(S^2)$ are the metric and volume 2-form of  round 2-sphere with unit radius, respectively, $g(\bR^6)$  is the Euclidean  metric
on $\bR^6$ and $W$ is a constant non-degenerate 2-form on  $\bR^6$. 

To continue let us focus on the gravitino KSE on $\sigma_+$. The integrability condition of this equation along the $\bR^6$ directions implies
that $W=\gamma \lambda$ and that $\Gamma_{1234}\sigma_+=\eta \sigma_+$ and $\Gamma_{1256}\sigma_+=\zeta \sigma_+$, where $\lambda$ is a K\"ahler form of the Euclidean metric on $\bR^6$ and we have chosen $\bR^6$ along the $123456$ directions. Without loss of generality one can always choose  $\lambda=\lambda_1 dx^1\wedge dx^2+ \lambda_2 dx^3\wedge dx^4+\lambda_3 dx^5\wedge dx^6$
with $\lambda_1, \lambda_2,\lambda_3=\pm 1$.  Then the algebraic KSE,  $\Xi^+\sigma_+=\big(-(2\ell)^{-1}+(288)^{-1} \Gamma_z {\slashed X}\big)\sigma_+=0$,  implies, after imposing $\Gamma_z\Gamma_{12} \Gamma_{(2)}\sigma_+=\theta \sigma_+$, $\theta=\pm1$,
that either $\gamma=\pm 6 \ell^{-1}$ or $\gamma=\pm 2 \ell^{-1}$, where $\Gamma_{(2)}$ is the product of two gamma matrices along two orthonormal directions tangential to $S^2$. The warp factor field equation is not satisfied for $\gamma=\pm 6 \ell^{-1}$. While for $\gamma=\pm 2 \ell^{-1}$, the Einstein equation along $S^2$ gives
$\ell=2$. This is the solution that describes the near horizon geometry of three intersecting M5-branes and preserves $1/4$ of supersymmetry.

After imposing all the conditions above on $\sigma_+$ appropriate for this solution, the gravitino KSE along $S^2$  can be written as
\bea
{\cal D}_\alpha^+=\nabla_\alpha\pm {1\over2}  \epsilon_{\alpha\beta} \Gamma^\beta \Gamma_{12}~,
\eea
where $a,b$ are restricted along two orthonormal tangential directions of $S^2$. Next it is straightforward to show that the 1-form bilinears restricted on $S^2$ are KY forms the 2-form bilinears restricted on $S^2$ are CCKY forms. So the 1-form bilinears generate symmetries for the probe action (\ref{espm}).

\section{Concluding remarks}

We have presented the TCFH of all supersymetric  AdS backgrounds of 11-dimensional supergravity. Therefore we have  demonstrated that all the form bilinears  on the internal space of these backgrounds are CKY forms with respect to the TCFH connection. This provides a geometric interpretation for all the conditions on these form bilinears implied by the KSE of the theory. We have also given the reduced holonomy of the TCFH connections for generic supersymmetric backgrounds and we have found that it factorises on the space symmetric and skew-symmetric form bilinears under the exchange of the two Killing spinors. We have illustrated our results with some examples that include the maximally supersymmetric AdS backgrounds of 11-dimensional supergravity as well as some other AdS backgrounds that arise as near horizon geometries of intersecting M-branes.
We have found  that some of the form bilinears on these backgrounds are KY forms and so generate symmetries for spinning particle probes
propagating on the internal spaces.

It is clear from the results of \cite{tcfhgp} that  TCFHs can be constructed for all supersymmetric theories that exhibit a gravitino KSE. The generality of this result allows for the exploration of many more possibilities. As a consequence, all form bilinears are CKY forms with respect to some connection. This indicates they may be related to symmetries for particle and string probes propagating on supersymmetric backgrounds.  Of course for special backgrounds this can be confirmed with  explicit calculations like those presented in section \ref{prob} of this paper. However given a TCFH of a supersymmetric theory, there is not  a systematic way to construct a probe action which exhibits symmetries generated by the form bilinears. Although there are many partial results that corroborate such constructions, the general question remains open.

\section*{Acknowledgments}

EPB is supported by the CONACYT, the Mexican Council of Science.

\end{document}